\documentclass{jaa}
\usepackage{natbib}
\usepackage{hyperref}
\hypersetup{colorlinks=true, linkcolor=blue, filecolor=blue, citecolor = blue, urlcolor=cyan,} 
\usepackage{graphicx}
\usepackage{float}
\usepackage{adjustbox} 
\usepackage{subcaption} 
\newcommand{\ie}{$i.e.,\;$}
\newcommand{\eg}{$e.g.,\;$}

\begin{document}\sloppy

\title{Characteristics of remnant radio galaxies detected in the deep radio continuum observations from the SKA pathfinders}


\author{Sushant Dutta\textsuperscript{1,2,*}, Veeresh Singh\textsuperscript{1}, C. H. Ishwara Chandra\textsuperscript{3}, Yogesh Wadadekar\textsuperscript{3} and Abhijit Kayal\textsuperscript{1,2}}
\affilOne{\textsuperscript{1}Physical Research Laboratory, Ahmedabad, 380009, Gujarat, India.\\}
\affilTwo{\textsuperscript{2}Indian Institue of Technology Gandhinagar, Palaj, Gandhinagar, 382355, Gujarat, India.\\}
\affilThree{\textsuperscript{3}National Centre for Radio Astrophysics, TIFR, Post Bag 3, Ganeshkhind,Pune 411007, India.\\}


\twocolumn[{

\maketitle

\corres{sushantd@prl.res.in}

\msinfo{27 April 2022}{15 July 2022}

\begin{abstract}
The cessation of AGN activity in radio galaxies leads to a remnant phase during which jets are no longer sustained,  but lobes can be detected for a period of time before they fade away due to radiative and dynamical energy losses. 
The time-scale of the remnant phase and AGN duty cycle are vital to understanding 
the evolution of radio galaxies. 
In this paper, we report new band-3 observations with the upgraded Giant Meterwave Radio Telescope (uGMRT) for five 
remnant radio galaxies.  
Our uGMRT observations reveal emission of low-surface-brightness in all five remnants with 400 MHz surface brightness in 
the range of 36$-$201 mJy~arcmin$^{-2}$. With band-3 uGMRT observations, we discover wing-shaped radio morphology in 
one of our sample sources. 
Using radio observations at 150 MHz, 325 MHz, 400 MHz, and 1.5 GHz, we model the radio spectral energy distributions 
(SEDs) of our sample sources with the continuous injection-off model (CI$_{\rm OFF}$), that assumes an active 
phase with continuous injection followed by a remnant phase. We obtain total source ages ($t_{\rm s}$) in 
the range of 20.3 Myr to 41.4 Myr with $t_{\rm OFF}$/$t_{\rm s}$ distributed in the range of 0.16 to 0.63, which in turn suggests them to belong to different evolutionary phases. 
We note that, in comparison to the remnants reported in the literature, 
our sample sources tend to show lower spectral ages that can be explained by the combined effects of 
more dominant inverse Compton losses for our sources present at the relatively higher redshifts 
and possible rapid expansion of lobes in their less dense environments.       
\end{abstract}

}]

%
%
\volnum{43}
\year{2022}
\pgrange{1--}
\setcounter{page}{1}
\lp{1}

\section{Introduction} 
\label{sec:intro}
Radio galaxies, a subclass of active galactic nuclei (AGN), exhibit highly collimated bipolar jets supplying plasma to radio lobes at large distances of hundreds of kilo-parsecs \citep[see][]{Hardcastle20,Saikia22}. 
The radio jets launched from the vicinity of an accreting super-massive black hole (SMBH) offer a direct probe to the status of AGN activity.  
The multi-frequency radio observations have demonstrated that the AGN activity in galaxies is only a phase (`active phase') lasting for a few tens of million years \citep{Dwarakanath09,Parma07,Saripalli12}. 
After the cessation of AGN activity, the outflowing jets are no longer 
sustained, which results in the stoppage of plasma supply to radio lobes. 
Therefore, the cessation of AGN activity in radio galaxies leads to a dying or remnant phase during 
which radio lobes can still be detected for a relatively short period of time before they fade away due 
to energy losses via radiative and dynamical processes \citep{Murgia11}. 
The small samples of remnants \citep{Mahatma18,Jurlin21}  
and the serendipitous discovery of individual remnant radio galaxies (hereafter remnants),  
{\eg}B2 0924+30 \citep{Cordey87,Shulevski17}, J021659-044920 \citep{Tamhane15}, blob1 \citep{Brienza16} and NGC 1534 \citep{Duchesne19}, WISEA J152228.01+274141.3 \citep{Lal21} etc., mainly detected in the low-frequency observations, led to the notion of remnants being rare sources. 
Remnants are also detected in the radio sources classified as the double-double radio galaxies (DDRGs) 
exhibiting two pairs of lobes resulting from the two different episodes of AGN-jet activity. 
In DDRGs, a new pair of inner lobes driven by the current phase of AGN-jet activity appears before the disappearance of the old pair of outer lobes representing the remnant phase of the previous episode of AGN-jet activity 
(\citet{Saikia09,Nandi12,Morganti17} and references therein). In our study, we consider remnant sources with no apparent episodic AGN-jet activity. 
Notably, such remnants seem to be even rarer than the DDRGs \citep{Mahatma19,Jurlin20}.
The rarity of remnants has conventionally been attributed to the rapid energy losses at work in relic lobes. 
Albeit, the timescale over which a remnant can be detected, varies from source to source 
and depends on various factors such as the source size, magnetic field in the lobes and the large-scale environment \citep{Turner18}.    
The evolution of radio galaxies in their remnant phase is not well understood due to their paucity.
\par
Also, with an aim to constrain the AGN duty cycle and evolutionary models, there have been attempts to perform systematic 
searches for the remnants \citep[see][]{Godfrey17,Brienza17,Mahatma18}. 
However, such studies often face limitations imposed by the selection criteria 
and the flux density limits of radio observations. 
For instance, using Ultra-Steep Spectrum (USS; ${\alpha}_{\rm 74~MHz}^{\rm 1.4~GHz}$ $<$ -1.2, where 
S$\nu$ $\propto$ ${\nu}^{\alpha}$) selection criterion \cite{Godfrey17} found 
that only $\leq$2$\%$ of their sample sources are remnant candidates.  
\cite{Brienza17} argued that the use of a single criterion is likely to miss a fraction of remnants, and both morphological as well as spectral criteria ought to be used for the identification of full remnant population. 
\cite{Mahatma18} emphasized that the absence of a radio core in deep radio observations can be considered 
as a reliable selection criterion for remnants, irrespective of their radio morphology and spectral behaviour. 
In their sample of 127 sources with S$_{\rm 150~MHz}$ $\geq$ 80~mJy and radio size $\geq$40$^{\prime\prime}$, \cite{Mahatma18} found only 11 remnants with no detected radio core in deep (noise-rms $\sim$ 0.02 mJy~beam$^{-1}$) 
6 GHz observations. 
More recently, \cite{Jurlin21} noted that a feeble radio core representing a dying AGN activity can still be present 
in a remnant and identified only 13 remnants in their sample of 158 sources. 
Notably, despite the concerted efforts, the studies based on the sensitive low-frequency radio observations in the deep fields, have yielded only small samples with remnant fraction limited to 5\%$-$10\% \citep[see][]{Mahatma18,Quici21,Jurlin21}. 
\par
Further, remnants selected using morphological criteria are of large angular sizes 
{\ie}$\geq$30$^{\prime\prime}$, where a limit on the minimum size is placed for deciphering the morphological details.  
Hence, morphologically selected remnants are inherently biased towards the remnants of more powerful large-size radio galaxies. 
Using the spectral curvature criterion, \cite{Singh21} identified remnants of small angular sizes ($<$30$^{\prime\prime}$) and 
found them to be more abundant, which is consistent with the theoretical models. 
For instance, using the `Radio AGNs in Semi-Analytic Environments' (RAiSE) dynamical model, 
\cite{Shabala20} showed that the models assuming a power law distribution of jet kinetic powers 
and source ages can explain the observed fraction of remnants and their distributions of flux density, 
angular size and redshift. 
In fact, the power law distribution of jet kinetic powers predicts more number of small-size remnants.    
Thus, the remnant population appears more diverse than that envisaged from the small samples of 
large-size remnants discovered earlier.  
Hence, it becomes important to study individual remnant sources and estimate various source parameters such as  
magnetic field, remnant age, AGN duty cycle and their large-scale environments. 
The detailed study of individual remnant sources can help us to understand the evolutionary scenario at work in different sources. 
In this paper, we study five remnant sources using multi-frequency radio observations. 
\par
This paper is organized as follows. In Section~\ref{sec:data} we provide the details of radio 
observations. The sample selection criteria and source characteristics are described in Section~\ref{sec:sample}. 
In Section~\ref{sec:energetic} we describe the radio SED modelling, spectral age determination of our sample remnants, 
and their comparison with the remnants reported in the literature. 
In Section~\ref{sec:SKA} we discuss the potential of deep and wide radio continuum 
radio surveys planned with the Square Kilometer Array (SKA) and its pathfinders. 
The results and conclusions of our study are presented in Section~\ref{sec:conclusions}.
\\ 
In this paper, we use the cosmological parameters H$_{\rm 0}$ = 70 km s$^{-1}$ Mpc$^{-1}$, ${\Omega}_{\rm m}$ = 0.3, and ${\Omega}_{\Lambda}$ = 0.7. Radio spectral index ${\alpha}$ is defined by assuming power law spectrum 
(S$_{\nu}$ $\propto$ ${\nu}^{\alpha}$, where S$_{\nu}$ denotes the flux density at frequency $\nu$).

\section{Radio observations} 
\label{sec:data}
For our study, we used band-3 observations from the upgraded Giant Metrewave Radio Telescope 
\citep[uGMRT,][]{Gupta17}, along with the auxiliary radio observations available at 325 MHz from the legacy 
Giant Metrewave Radio Telescope \citep[GMRT,][]{Swarup91}, 
144 MHz observations from the LOw Frequency ARray \citep[LOFAR,][]{VanHaarlem13} and 1.5 GHz observations from the Jansky Very Large Array \citep[JVLA][]{Perley11}. 
All our remnant sources are located within a small region of 2.3 deg$^{2}$ of 
the {\em XMM}$-$Large Scale Structure ({\em XMM-LSS}) field. 
In Table~\ref{tab:Data}, we list the basic details of the radio observations.  
We give a brief description of these observations in the following subsections. 
\begin{table}[ht]
\centering
\caption{Summary of radio observations}
\label{tab:Data}
\scalebox{0.85}{
\begin{tabular}{ccccc}
\topline
Telescope   & band    & ${\nu}_{\rm central}$      &  noise-rms      & beam-size    \\  
          &  (MHz)  &  (MHz)      & (mJy~beam$^{-1}$) & ($^{\prime\prime}$ $\times$ $^{\prime\prime}$)  \\ \midline
LOFAR       & 120$-$168   & 144   &  0.28$-$0.39   & 7.5$\times$8.5    \\
GMRT        & 309$-$341   &  325  &  0.15          & 10.2$\times$7.9    \\
uGMRT       & 250$-$500   &  400  &  0.03          & 6.7$\times$5.3     \\
JVLA        & 994$-$2018  &  1500 &  0.016         & 4.5$\times$4.5     \\
\hline
\end{tabular}}
\end{table}

\subsection{Band-3 uGMRT radio observations}
Our band-3 (250 MHz $-$ 550 MHz) uGMRT observations were performed on 11 September 2017 
(proposal code 32$\_$066; PI: C.H. Ishwara-Chandra) in 
full synthesis mode with nearly 10 hours of total observing time. To obtain the deepest possible image, 
these observations were limited only to one pointing centered at RA = 02$^{\rm h}$ 26$^{\rm m}$ 45$^{\rm s}$.0 and DEC = -04$^{\circ}$ 41$^{\prime}$ 30$^{\prime\prime}$.0 in the {\em XMM-LSS} field. 
The observations were acquired in a standard manner by observing 
the flux calibrators 3C48 and 
3C147 for 15$-$20 minutes at the beginning and end of the observing session, respectively. The phase calibrators 
0116-0208 and 0323+055 were observed for nearly 5 minutes with scans inter-leaved with 30 minutes scans of the targeted field. 
The uGMRT radio data were reduced using a pipeline based on the Common Astronomy Software
Applications (CASA\footnote{https://casa.nrao.edu/}) routines. We followed the standard data reduction 
procedure that involved flagging bad data, calibrating visibilities, imaging of calibrated visibilities, and several iterations of self-calibration.     
The final primary beam corrected image achieves a median noise-rms of 30~$\mu$Jy~beam$^{-1}$. 
As expected, the noise-rms is lowest ($\sim$20~$\mu$Jy~beam$^{-1}$) in the central region, while a higher noise-rms 
($\sim$40~$\mu$Jy~beam$^{-1}$) is seen around bright sources and in the peripheral regions. 
The synthesized beam-size is 6$^{\prime\prime}$.7 $\times$ 5$^{\prime\prime}$.3.       
Using the Python Blob Detector and Source Finder (PyBDSF; \citet{Mohan15}) algorithm, we detected a total number of 
2332 sources at $\geq$5$\sigma$ over the full sky-area of 2.3 deg$^{2}$. 
More details on the data reduction and creation of the source catalogue can be found in a companion paper 
(Kayal et al. 2022, JoAA, {\textit {accepted}}). 
\subsection{Auxiliary radio observations at 144 MHz, 325 MHz and 1.5 GHz}
To build radio spectral energy distributions (SEDs) of our remnants, we utilized radio observations 
available at other frequencies over the widest range. 
We used 144 MHz LOFAR observations available in the {\em XMM-LSS} field \citep{Hale19}. 
The LOFAR observations centered at RA = 35$^{\circ}$ and Dec = -4.5$^{\circ}$ cover $\sim$27 deg$^2$ sky-area. 
The final mosaiced image achieves noise-rms down to 0.28 mJy beam$^{-1}$ in the central regions while median noise-rms is 
0.394 mJy~beam$^{-1}$. The LOFAR observations have a resolution 
(7$^{\prime\prime}$.5 $\times$ 8$^{\prime\prime}$.5), similar to our band-3 uGMRT observations. 
\par
The region of our band-3 uGMRT observations is also covered with the 325 MHz observations performed with 
the legacy GMRT equipped with software correlator giving 32 MHz instantaneous bandwidth (Proposal: 20\_006; PI: Y. Wadadekar). 
The mosaiced 325 MHz image centred at RA = 02$^{\rm h}$ 21$^{\rm m}$ 00$^{\rm s}$ and DEC = -04$^{\circ}$ 30$^{\prime}$ 
00$^{\prime\prime}$ covers 12.5 deg$^2$ sky-area in the {\em XMM-LSS} field. 
The 325 MHz mosaiced image has an average noise-rms of 150 $\mu$Jy beam$^{-1}$ and synthesized beam-size of 
10$^{\prime\prime}$.2 $\times$ 7$^{\prime\prime}$.9. More details about 325 MHz GMRT observations can be 
found in \cite{Singh14}. 
\par
We also utilized 1.5 GHz JVLA observations reported by \cite{Heywood20}. 
The 1.5 GHz observations performed with the VLA in B-configuration using wide-band (0.994-2.018 GHz) continuum mode 
cover 5.0 deg$^2$ sky-area overlapping with the NIR VIDEO region in the {\em XMM-LSS} field \citep{Jarvis13}. 
The mosaiced image has nearly uniform noise-rms with a median value of $\sim$ 16 $\mu$Jy beam$^{-1}$ 
and synthesized beam-size of 4$^{\prime\prime}{.}$5 $\times$ 4$^{\prime\prime}{.}$5. 
The JVLA observations with the better resolution are efficient in 
revealing structural details. Although, while building radio SEDs we ensure that the JVLA observations do not suffer 
with the missing flux issue wherein diffuse emission of low-surface-brightness can be missed due to 
its higher resolution.   
We compare JVLA flux densities with those from the 1.4 GHz NVSS, whenever available, and we use NVSS flux 
densities in case of the missing flux issue. 
With its large beam-size of 45$^{\prime\prime}$ NVSS observations can efficiently detect the diffuse emission 
of low-surface-brightness emission. 
\par 
We note that the sky region of our band-3 uGMRT observations is also covered with the 240 MHz, 610 MHz GMRT observations \citep{Tasse07} and 74 MHz VLA observations \citep{Tasse06}. In principle, these observations can be 
useful in building radio SEDs, however, we found that, unexpectedly, 240 MHz and 610 MHz flux densities are systematically 
underestimated \cite[see][]{Singh21}. Hence, we prefer not to use 240 MHz and 610 MHz GMRT observations. Also, 74 MHz VLA observations yielding noise-rms of 32 mJy~beam$^{-1}$ are too shallow to detect our relatively faint sources. 
Therefore, in our study, we used radio observations mainly from the LOFAR at 144 MHz, the GMRT at 325 MHz, the uGMRT at 400 MHz, 
and the JVLA at 1.5~GHz.     
\begin{table*}
\caption{Remnant sample}
\label{tab:sample}
\scalebox{0.75}{
\begin{tabular}{ccccccccccc}
\hline
Source & S$^{\rm int}_{\rm 150~MHz}$ & S$^{\rm int}_{\rm 325~MHz}$ & S$^{\rm int}_{\rm 400~MHz}$ & S$^{\rm int}_{\rm 1.4~GHz}$ & ${\alpha}_{\rm 150~MHz}^{\rm 325~MHz}$ & ${\alpha}_{\rm 400~MHz}^{\rm 1.5~GHz}$ & SB$_{\rm 400~MHz}$ &  $z$   & Size &  logL$_{\rm 400~MHz}$\\
Name & (mJy) & (mJy) & (mJy) & (mJy)  &        &       &  (mJy~arcmin$^{-2}$) &    &  (kpc($^{\prime \prime}$))   &       \\
\hline
GMRT022318-044526 & 33.9$\pm$1.0 & 20.2$\pm$2.2 & 12.1$\pm$1.4 & 1.70$\pm$0.10 & -0.67$\pm$0.15 & -1.48$\pm$0.10 & 106  & 1.15$\pm$0.17 & 643 (78)  & 25.99 \\
GMRT022338-045418 & 7.04$\pm$0.3 & 3.7$\pm$0.7  & 3.1$\pm$0.7 & 0.6$\pm$0.09   & -0.83$\pm$0.25 & -1.24$\pm$0.21 & 36 & 0.81$\pm$0.01 & 200 (26.5) & 25.29  \\
GMRT022723-051242 & 45.7$\pm$0.8 & 21.1$\pm$1.1 & 15.8$\pm$0.5 & 2.20$\pm$0.30 & -0.98$\pm$0.07 & -1.49$\pm$0.11 & 201 & 0.87$\pm$0.07 & 255 (33) & 26.06 \\
GMRT022737-052139 & 6.3$\pm$0.8 & 3.7$\pm$0.60 & 3.3$\pm$0.3 & 0.6$\pm$0.07  & -0.69$\pm$0.27 & -1.29$\pm$0.11 & 56  & 0.62$\pm$0.13 & 190 (28) & 25.25  \\
GMRT022802-041417 & 37.3$\pm$0.9 & 19.5$\pm$1.0 & 16.2$\pm$0.5 & 2.4$\pm$0.2 & -0.84$\pm$0.07 & -1.44$\pm$0.07 & 127 & 0.44$\pm$0.05 & 216 (38) & 25.19 \\
\hline
\end{tabular}}	
\end{table*}
\section{The remnant sample and band-3 uGMRT observations}
\label{sec:sample}
In this subsection, we describe the sample selection criteria and the characteristics of our remnant 
sources.

\subsection{Remnant sample and selection criteria}
Our five remnant sources studied in this paper are taken from \cite{Singh21} and Dutta et al. 2022 ({\it under review}). 
We list the basic parameters, {\ie}flux densities, surface brightness, radio sizes, redshifts, and 400 MHz radio 
luminosities of our sample sources in Table~\ref{tab:sample}. 
All the five remnant sources are selected by using the spectral curvature criterion,  
{\ie}${\alpha}_{\rm 150~MHz}^{\rm 325~MHz}$ - ${\alpha}_{\rm 325~MHz}^{\rm 1.5~GHz}$ 
$\geq$0.5.  
Using 150 MHz LOFAR, 325 MHz GMRT and 1.4 GHz JVLA observations \cite{Singh21} demonstrated 
that the remnant sources of small angular sizes ($<$30$^{\prime\prime}$) often lacking morphological details in the radio images 
of 5$^{\prime\prime}$ $-$ 10$^{\prime\prime}$ angular resolution, can be identified by using spectral curvature criterion 
(${\alpha}_{\rm low}$ - ${\alpha}_{\rm high}$ $\geq$0.5). 
We note that our five remnant sources also satisfy morphological criteria, 
{\ie}diffuse emission of low-surface-brightness (36$-$201 mJy~arcmin$^{-2}$ at 400 MHz, see Table~\ref{tab:sample}) 
and an absence of core in the high frequency 3.0 GHz VLA Sky Survey \citep[VLASS;][]{Lacy20} images. 
The morphological criteria alone can enable us to make robust identification of remnants irrespective of their 
spectral characteristics \citep{Hota11,Mahatma18}. Therefore, our sources satisfying both morphological and 
spectral criteria can be regarded as the confirmed remnants. 
\par
Further, we note that unlike most of the remnant sources reported in the literature, our sample sources are 
relatively faint (150 MHz flux density in the range of 6.3 mJy to 37.3 mJy) and reside at relatively 
higher redshifts ($z$ $\geq$ 0.44) (see Table~\ref{tab:sample}). 
It is worth mentioning that most of the individually studied remnant sources 
\citep[{\eg}][]{Brienza16,Shulevski17,Duchesne19} as well as the remnants derived from the LOFAR observations 
are relatively bright, {\ie}S$_{\rm 150~MHz}$ $>$80~mJy and extended 
($\geq$40$^{\prime\prime}$) \citep[see][]{Mahatma18,Jurlin21}. Therefore, our sample of relatively faint and 
small-size remnants allow us to probe a different phase space.  

\subsection{Characteristics of our remnants}
In the following subsections, we describe radio morphologies and characteristics of our sample sources.
Our new 400 MHz uGMRT observations discover the wing-shaped radio morphology in GMRT022338-045418 
(see~Figure~\ref{fig:uGMRTz}). To compare the radio morphological details at different frequencies, we plot 
radio contours of 400~MHz uGMRT, 325 MHz GMRT and 1.5 GHz JVLA emissions onto 
the corresponding optical $i$ band images from the Hyper Suprime-Cam Subaru Strategic Program 
(HSC-SSP\footnote{https://hsc.mtk.nao.ac.jp/ssp/}) (see Figure~\ref{fig:ImSED}, {\it left panel}). 

%
%
\subsubsection{GMRT022318-044526:} 
This source shows a clear double-lobed radio morphology with a total end-to-end projected size of 78$^{\prime\prime}$ 
that corresponds to the physical size of 643 kpc at the redshift $z$ = 1.15$\pm$0.17 of the potential host. 
The deep 1.5 GHz JVLA observations detect no core but only two lobes that appear edge-brightened 
similar to the FR-II radio galaxies. 
Our new 400 MHz uGMRT observations revealed nearly the same structures that were seen in 
the previous 325 MHz observations (see Figure~\ref{fig:ImSED}, {\it top left panel}). 
Based on the radio contours overlaid on the HSC-SSP $i$ band image, a galaxy J022318.1-044520.7 matching close to the 
centroid of the low-frequency radio emission, is identified as the potential host with an $i$ band 
magnitude ($m_{i}$) 24.09$\pm$0.08 and photometric redshift ($z_{\rm phot}$) 1.15$\pm$0.17. 
\subsubsection{GMRT022338-045418:}
\begin{figure}[ht]
\includegraphics[angle=0,width=8.0cm,trim={1.0cm 3.25cm 7.5cm 3.25cm},clip]{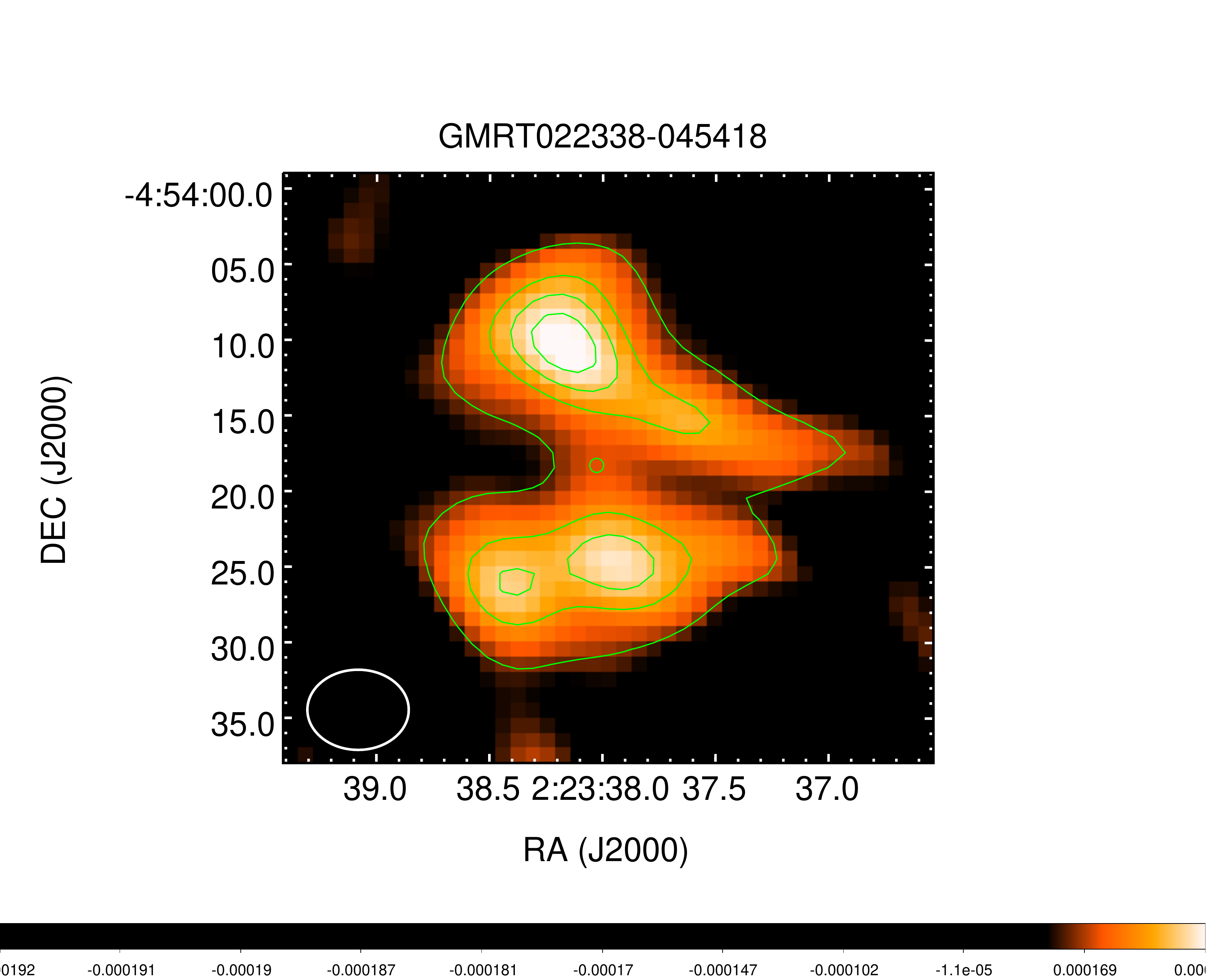}
\caption{The band-3 uGMRT image of remnant source GMRT022338-045418. For a better representation of 
intensity distribution, radio contours (in green colour) are overplotted onto the corresponding false colour radio image. 
The radio contours are shown at $\sigma$~$\times$~(3, 6, 10, 15......) levels. The centroid position is marked 
with a small green circle. The uGMRT beam-size of 6$^{\prime\prime}$.7~$\times$~5$^{\prime\prime}$.3 is shown 
in the bottom left corner.}
\label{fig:uGMRTz}
\end{figure}

The band-3 uGMRT image of GMRT022338-045418 shows a wing-shaped radio morphology 
(see Figure~\ref{fig:uGMRTz}),  
wherein the northern (upper) component resembles a back-flow-like feature, while the southern (lower) component 
shows a double-peaked emission. 
This source has very low surface-brightness emission (36 mJy arcmin$^{-2}$ at 400 MHz) such that the southern component 
is completely undetected in the 1.5 GHz JVLA image, while 325 MHz GMRT image of relatively low resolution 
(10$^{\prime\prime}$.2~$\times$~7$^{\prime\prime}$.9) shows only north-south elongation with no apparent structures 
(see Figure~\ref{fig:ImSED}, {\it middle left panel}). 
Interestingly, the northern component which resembles with a backflow tail in the 400 MHz uGMRT 
image clearly shows two peaks in the 1.5 GHz JVLA 
image (see Figure~\ref{fig:ImSED}, {\it middle left panel}). 
Thus, we find that both northern and southern components have two-peak emission, which 
is commonly seen in the winged radio galaxies classified as the $Z$-shaped and $X$-shaped radio sources, wherein diagonally opposite bright peaks correspond to primary lobes, and fainter peaks correspond to secondary lobes, possibly from the previous episode of AGN jet activity \citep{Lal19,Bera22}.       
\par
We note that the discovery of wing-shaped radio morphology in GMRT022338-045418 is credited to the sensitivity 
and resolution 
of band-3 uGMRT observations that are adequate to detect faint diffuse radio emissions. 
The total flux density at 400 MHz of this source is only 3.1$\pm$0.5 mJy, while northern and southern components 
have a flux density of 1.62$\pm$0.3 mJy and 1.48$\pm$0.3 mJy, respectively.
The lack of its complete detection in all but band-3 observations poses a challenge in characterizing this source. 
For instance, a spectral index map requires images at more than one widely separated frequencies with similar 
angular resolutions. Also, with the present set of radio observations we cannot rule out if wing-shaped radio morphology 
is the result of two individual double$-$lobe radio sources located close to each other. 
Deeper radio observations of higher resolution are needed to examine if this source consists of 
two individual radio sources.  
Although, given the lack of optical counterparts matching with individual lobes and a backflow-like feature seen in the band-3 
image (see Figure~\ref{fig:ImSED}, {\it left panel}) suggest that the wing-shaped morphology is unlikely to be 
the result of two individual double$-$lobed radio sources coincidentally placed adjacent to each other. 
We note that there is no detected optical counterpart close to the centroid of the band-3 radio morphology 
(see Figure~\ref{fig:ImSED}). 
However, within a radius of 10$^{\prime\prime}$ there are two optical sources, {\ie}a relatively bright source 
J022337.7-045414.8 towards the north-west with $m_{\rm i}$ = 22.29$\pm$0.01 and photometric 
redshift ($z_{\rm phot}$) 0.81$\pm$0.01, and a faint source J022338.2-045418.6 towards the east 
with $m_{\rm i}$ = 24.99$\pm$0.05 and photometric redshift ($z_{\rm phot}$) 2.75$\pm$0.56. 
Because radio galaxies generally host in bright ellipticals, we assumed the brighter optical 
counterpart J022337.7-045414.8 as the potential host.
\subsubsection{GMRT022723-051242:} The radio morphology of this source resembles a double-lobed source, but the two lobes are not well 
resolved due to its small angular size of 33$^{\prime\prime}$. 
We note that a separate point radio source J022723.2-051223.8 is located close to its northern side 
(see Figure~\ref{fig:ImSED}, {\it left panel}). 
Due to its proximity, the point radio source contaminates the emission of the remnant source 
at 150 MHz, 325 MHz and 400 MHz as the spatial resolution is somewhat coarser at these frequencies. 
To accurately determine the flux density of our remnant 
source, we removed the contamination of the point radio source by considering its 1.5 GHz 
flux density from the JVLA image and extrapolating it to the lower frequencies by assuming a spectral index of -0.7.     
The 1.5 GHz JVLA radio contours of our remnant show two emission peaks each apparently corresponding to 
the northern and southern lobes, respectively. 
It is worth noting that the southern peak coincides with a foreground galaxy 
(J022722.9-051250.6) having $m_{i}$ = 19.90$\pm$0.05 and 
$z_{\rm phot}$ = 0.26$\pm$0.10. From the current observations, it is unclear if the foreground galaxy also contaminates and to what extent. Although considering the relaxed radio morphology of our remnant with an evident 
diffusion of emission along the direction perpendicular to the jet-axis, 
the foreground galaxy is unlikely to give rise to any significant contamination. 
Also, the ultra-steep radio spectrum towards high-frequencies (${\alpha}_{\rm 400~MHz}^{\rm 1.5~GHz}$ = -1.49$\pm$0.11) 
suggests only minimal contamination from the foreground galaxy, if it exists. 
Based on the radio contours, in particular from the 1.5 GHz JVLA overlaid on the $i$ band optical image, 
we consider J022723.1-051242.9 as a potential host with $m_{i}$ = 23.02$\pm$0.01 and $z_{\rm phot}$ = 0.87$\pm$0.07. 
However, another galaxy J022722.9-051242.8, located close to its centroid, cannot be ruled out from being 
a plausible host. Our new uGMRT radio observations show radio morphology similar to that seen with the 325 MHz GMRT 
and 1.5 GHz JVLA observations.      
\subsubsection{GMRT022737-052139:}
The radio morphology of this source shows east-west elongation. The high resolution 1.5 GHz JVLA image indicates 
the double-lobe-like structure (see Figure~\ref{fig:ImSED}, {\it left panel}). The two lobes are not well resolved due 
to the small angular size (28$^{\prime\prime}$) of this source. 
Our new uGMRT observations show emission consistent with the 325 MHz GMRT observations. 
With radio contours overlaid on the $i$ band optical image we identify J022737-052139 as a potential host 
with $m_{i}$ = 24.17$\pm$0.03 with $z_{\rm phot}$ = 0.62$\pm$0.13. 
With a total size of 190 kpc this source is the smallest in our sample. The strong spectral curvature and 
low surface brightness of 56 mJy~arcsec$^{-2}$ at 400~MHz suggest it to be a remnant source.
\subsubsection{GMRT022802-041417:}
This source is hosted in a nearby bright ($m_{i}$ = 19.45$\pm$0.01) elliptical galaxy with a photometric 
redshift of 0.44$\pm$0.05. The angular size (38$^{\prime\prime}$) of this source corresponds to the 
total projected linear size of 216 kpc. Due to its relatively small angular size, 325 MHz and 400 MHz observations 
do not reveal clear structures. However, 1.5 GHz JVLA 
image showing east-west elongation and two tentative peaks infer it to be a double-lobed source 
(see Figure~\ref{fig:ImSED}, {\it left panel}).
\section{Spectral ages} 
\label{sec:energetic}
In this section, we estimate the spectral ages of our remnants and compare them with the remnants already studied in the literature.    
\subsection{SED modelling}
\label{sec:age}
We derive the spectral ages of our remnants via modelling their SEDs with physically motivated models. 
We note that the radio spectrum of an active radio galaxy can be characterized by the continuous injection model (CI model, 
\citet{Kardashev62,Jaffe73}) that assumes continuous replenishment of radio lobes at a constant 
rate for the duration of $t_{\rm ON}$. The relativistic particles emitting radio emission have power law energy 
distribution ($N(E)$ $\propto$ $E^{-p}$) which results in a power law radio spectrum with an index 
of ${\alpha}_{\rm inj}$ = ($p$ - 1)/2. Due to the energy-dependent cooling rate of relativistic particles, the 
radio spectrum develops a break during the active phase itself, such that the spectral index above 
${\nu}_{\rm b}$ steepens to the value of ${\alpha}_{\rm inj}$ $-$ 0.5, where ${\alpha}_{\rm inj}$ is typically found in the range of 
-0.5 to -0.8. As the radio source evolves, the break frequency (${\nu}_{\rm b}$) shifts progressively towards the lower frequencies.  
The relation between source age ($t_{\rm s}$) and break frequency ${\nu}_{\rm b}$ can be expressed as below 
\citep[see][]{Komissarov94,Slee01,Parma07}. 
\begin{table*}[ht]
\centering
\caption{Spectral ageing and source parameters}
\label{tab:energetics}
\begin{adjustbox}{width=\textwidth}
\begin{tabular}{cccccccccccccc}
\hline
Source & ${\alpha}_{\rm inj}$ & $d$ & $I_{\rm 0}$ & $\xi(\alpha, \nu_{1}, \nu_{2})$ & $u_{min}$ & B$_{\rm eq}$ & $\nu_{\rm b,~low}$ & $\nu_{\rm b,~high}$ & $t_{\rm s}$ & $t_{\rm ON}$ & $t_{\rm OFF}$  & $\frac{t_{\rm OFF}}{t_{\rm s}}$   & ${\chi}_{\rm red}^{2}$ \\
Name   &     & (kpc) & (mJy arcsec$^{-2}$) &      & (erg cm$^{-3}$) & ($\mu$G) & (GHz)    & (GHz)  & (Myr)  & (Myr)  & (Myr)  &    &   \\
\hline
GMRT022318-044526 & -0.40  & 188  & 0.029  &  3.79 $\times$ 10$^{-12}$ & 6.61 $\times$ 10$^{-13}$ & 2.66 & 0.255 & 3.60 & 20.3$^{+2.5}_{-2.8}$ & 15.2$^{+1.9}_{-1.7}$ & 5.1$^{+1.0}_{-1.2}$ &  0.25  &  1.94 \\
GMRT022338-045418 & -0.54  &  187  & 0.010  & 2.50 $\times$ 10$^{-12}$ &  1.24 $\times$ 10$^{-12}$  & 3.66 & 0.702 & 4.32 & 24.3$^{+2.7}_{-2.6}$ & 14.5$^{+1.7}_{-1.5}$ & 9.8$^{+1.9}_{-1.6}$ & 0.40  &  1.09  \\
GMRT022723-051242 & -0.62  & 197  & 0.056 & 1.72 $\times$ 10$^{-12}$ & 7.35 $\times$ 10$^{-13}$  & 2.80 & 0.251 & 10.9 & 33.2$^{+3.8}_{-3.6}$ & 28.0$^{+3.4}_{-3.1}$ & 5.2$^{+1.5}_{-2.8}$ &  0.16       & 1.08  \\
GMRT022737-052139 & -0.54  & 94  & 0.016  & 2.50 $\times$ 10$^{-12}$ & 4.37 $\times$ 10$^{-13}$ & 2.17  & 0.875 & 3.89  & 27.5$^{+4.3}_{-3.6}$ & 14.5$^{+4.2}_{-3.3}$ &  13.0$^{+1.1}_{-1.6}$  & 0.47 & 1.01   \\
GMRT022802-041417 & -0.56   & 157 & 0.035  & 1.72 $\times$ 10$^{-12}$ & 3.03 $\times$ 10$^{-13}$ & 1.80   & 0.812 & 2.05  & 41.4$^{+2.8}_{-2.2}$ & 15.4$^{+2.5}_{-1.9}$ & 26.0$^{+1.1}_{-1.0}$ &  0.63    & 1.08 \\
\hline
\end{tabular}
\end{adjustbox}	
\end{table*}
\begin{equation}
\label{eq:1}
t_{\rm s} = 1590 \Bigg[~\frac{B^{\rm 0.5}_{\rm eq}}{(B_{\rm eq}^{2} + B_{\rm CMB}^{2})\sqrt{{\nu}_{\rm b}(1 + z)}}~\Bigg]~{\rm Myr}
\end{equation}
where $t_{\rm s}$ is the total source age, $B_{\rm eq}$ and $B_{\rm CMB}$ $=$3.25(1+$z$)$^2$ are equipartition magnetic field and inverse Compton equivalent magnetic field, respectively, in the unit of $\mu$G, and $\nu_{\rm b}$ is the break frequency in GHz 
above which spectrum steepens from the initial injection spectral index. We note that the spectral age 
calculation assumes a uniform magnetic field strength across all the emitting regions, and it also assumes that the magnetic 
field remains constant during the radiative cooling process. 
Further, the model considers only radiative losses without accounting for the losses due to expansion.  
\par
To estimate the age of remnants, we use continuous injection off (CI$_{\rm OFF}$) model \citep{Komissarov94} that considers  
remnant phase after switching off the continuous injection phase lasting for the duration 
of $t_{\rm ON}$. In the remnant phase, a new break appears 
at higher frequencies (${\nu}_{\rm b,~high}$) beyond which spectrum drops exponentially. 
As the remnant phase progresses, ${\nu}_{\rm b,~high}$ shifts towards the lower frequencies. 
The ratio of $\nu_{\rm b,~low}$ and $\nu_{\rm b,~high}$ depends on the duration of remnant phase 
($t_{\rm OFF}$) {\it w.r.t.} the total source age as given by the following formula. 
\begin{equation}
\label{eq:2}
\frac{t_{\rm OFF}}{t_{\rm s}} = \Bigg(\frac{\nu_{\rm b,~low}}{\nu_{\rm b,~high}}\Bigg)^{0.5}
\end{equation}
where $t_{\rm s}$ is the total source age and can be expressed as $t_{\rm s}$ = $t_{\rm ON}$ + $t_{\rm OFF}$. 
It is obvious that ${\nu}_{\rm b,~low}$ would be lower than ${\nu}_{\rm b,~high}$ as $t_{\rm OFF}$ $<$ $t_{\rm s}$. 
However, for a very old remnant, ${\nu}_{\rm b,~high}$ would appear sufficiently close to ${\nu}_{\rm b,~low}$. 
We note that the computation of spectral age requires knowledge of magnetic field strength (see Eqn~\ref{eq:1}). 
We estimate magnetic field strength by assuming equipartition of 
minimum energy between particles and magnetic field and use the following equation.
\begin{equation}
\label{eq:3}
B_{\rm eq} [G] = \left(\frac{24 \pi}{7} u_{min}\right)^{1/2}
\end{equation}
where minimum particle energy ($u_{min}$) can be calculated using Equation~\ref{eq:4} with an approximation of isotropic 
particle distribution \citep[see][]{Govoni04}. 
\begin{equation}
\label{eq:4}
\begin{aligned}
u_{min}\left[\frac{\rm erg}{\rm cm^3}\right] = \xi(\alpha, \nu_{1}, \nu_{2})(1+k)^{4/7}(\nu_{0}[\rm MHz])^{-4\alpha/7}\times \\
(1+z)^{(12-4\alpha)/7}\left(I_{0}[\rm mJy~arcsec^{-2}]\right)^{4/7}(d[\rm kpc])^{-4/7}
\end{aligned} 
\end{equation}
where parameter $\xi$ depends on the spectral index ($\alpha$ $<$ 0 is considered for the observed spectrum) 
and spectrum is integrated over frequencies 
${\nu}_{\rm 1}$ and ${\nu}_{\rm 2}$ that correspond to 10 MHz and 100 GHz, respectively, 
which in turn correspond to a minimum Lorentz factor (${\gamma}_{\rm min}$) 10 and maximum Lorentz factor 
(${\gamma}_{\rm max}$) 10$^{5}$. We use $\xi$ values from the Table 1 of \cite{Govoni04}.
The parameter $k$ represents the energy ratio of relativistic protons to 
electrons and is assumed to be 1 for our calculations, $I_{0}$ is surface brightness at measuring frequency which is 400 MHz in our case, and $d$ is the source depth. 
The geometry of our sources is assumed to be cylindrical, and the cross-sectional diameter is taken as 
the source depth. 
\par
The estimated values of minimum particle 
energy ($u_{\rm min}$), equipartition magnetic field strength $B_{\rm eq}$ of our remnant sources are 
given in Table~\ref{tab:energetics}. 
We note that the $B_{\rm eq}$ values for our remnants are found in the range of 1.8~$\mu$G to 3.66~$\mu$G, which are similar to 
that reported for the remnants studied in the literature, 
{\eg}1.0~$\mu$G for BLOB1 \citep{Brienza16}, 0.89$-$1.6~$\mu$G for B2 0924+30 
\citep{Jamrozy04,Shulevski17}, 2.7~$\mu$G for NGC~1534 \citep{Duchesne19} and 4.1~$\mu$G for J1615+5452 \citep{Randriamanakoto20}. The remnants in DDRGs also show the magnetic field strength 
of a few $\mu$G \citep[see][]{Konar12}.   
To estimate the break frequencies (${\nu}_{\rm b,~low}$ and ${\nu}_{\rm b,~high}$) and spectral ages of our remnants, we model their radio spectra by using the Broad-band Radio Astronomy ToolS 
({\tt BRATS}\footnote{http://www.askanastronomer.co.uk/brats/}; \cite{Harwood13}) package. 
Given the remnant nature of our sources, we use CI$_{\rm OFF}$ model, also known as the KGJP model \citep{Komissarov94}, which is a modified form of the continuous injection 
model (CI; \cite{Jaffe73}) with the inclusion of energy losses via synchrotron and inverse Compton emission, 
after the cessation of AGN activity. 
While fitting the radio spectra in the {\tt BRATS} 
we provide input parameters $B_{\rm eq}$ and ${\alpha}_{\rm inj}$, and obtain output parameters such as 
the break frequencies (${\nu}_{\rm low}$ and ${\nu}_{\rm high}$), time-scales of active ($t_{\rm ON}$) and remnant 
($t_{\rm OFF}$) phases, total source age ($t_{\rm s}$), and the goodness of fit parameter reduced ${\chi}^{2}$. 
We point out that the accurate value of ${\alpha}_{\rm inj}$ is not known a priori, and hence, we begin with a value similar to 
${\alpha}_{\rm 150~MHz}^{\rm 1.5~GHz}$ which is the low frequency spectral index available for our sample sources. 
The best fit values of ${\alpha}_{\rm inj}$ for our sources are in the range -0.5 $-$ -0.6 which are similar to the 
typical values measured in remnant and active radio galaxies \citep{Murgia11}.  
Although, one of our sample sources GMRT022318-044526, the radio SED is best fitted with a somewhat lower 
value of ${\alpha}_{\rm inj}$ = -0.4 (see Table~\ref{tab:energetics}). 
\par
The modelled radio spectra of our sources are shown in Figure~\ref{fig:ImSED} ({\it right panel}), and the 
best-fit parameters are listed in Table~\ref{tab:energetics}. 
%
%
\begin{figure*}[hp]
\includegraphics[angle=0,width=8.0cm,trim={0.0cm 4.0cm 0.0cm 6.5cm},clip]{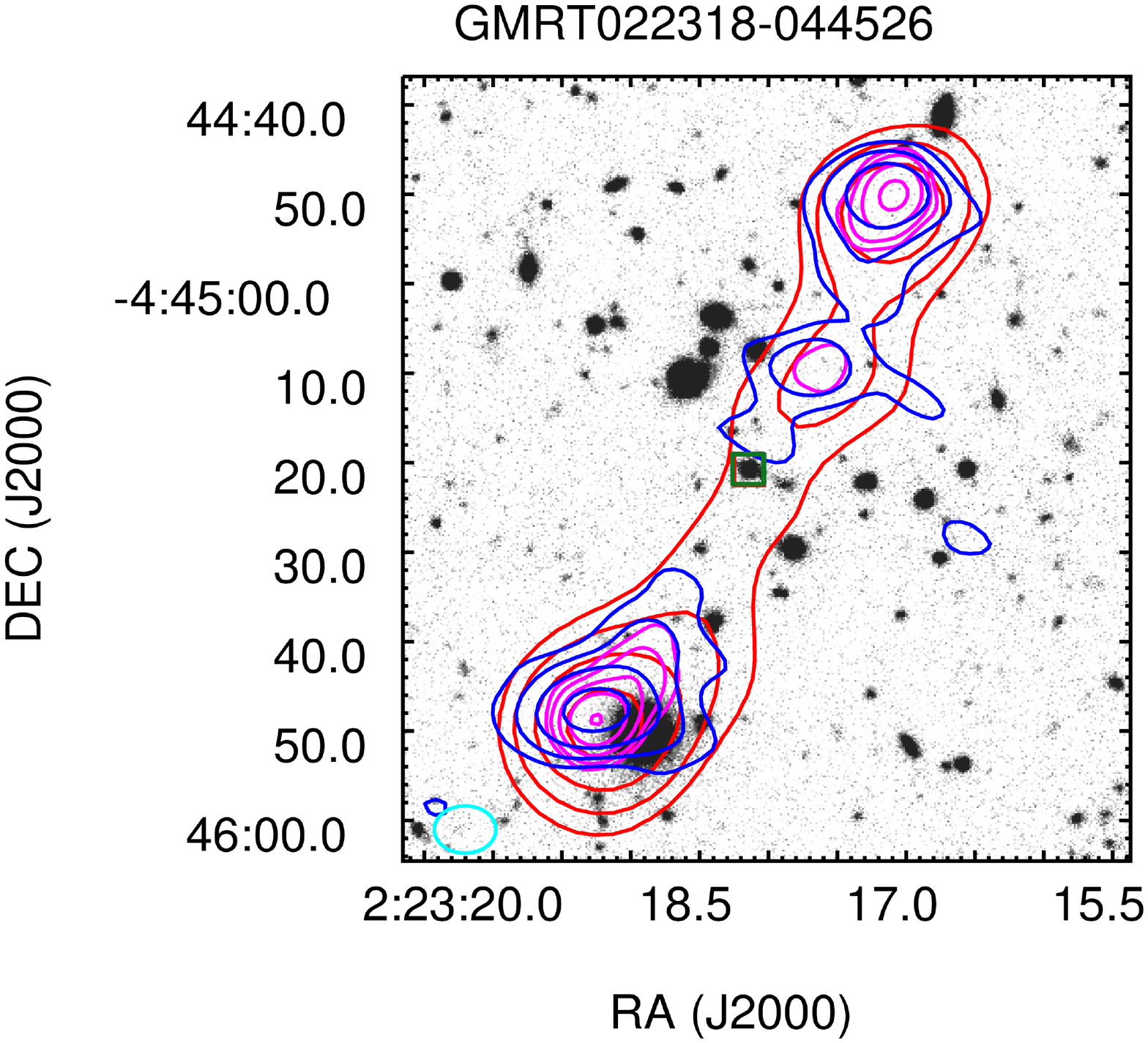}
\includegraphics[angle=0,width=9.0cm,trim={0.0cm 0.0cm 0.0cm 0.0cm},clip]{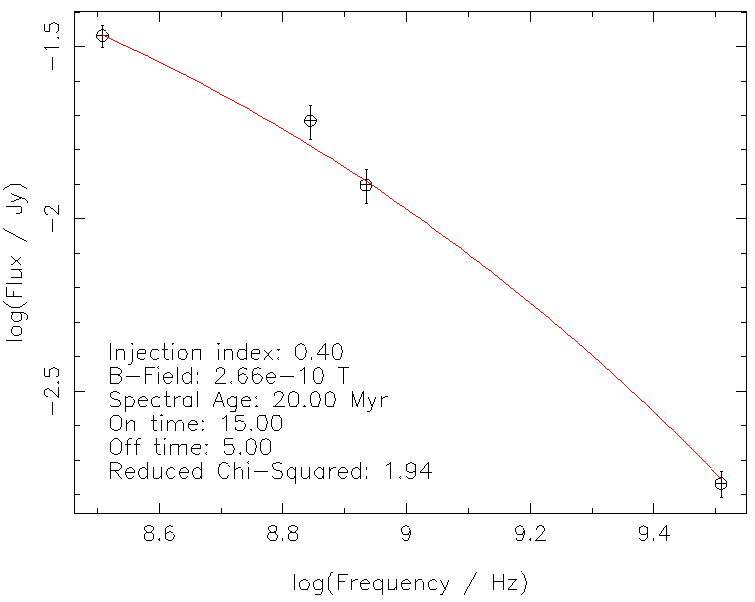}
\includegraphics[angle=0,width=8.0cm,trim={0.0cm 4.0cm 0.0cm 6.5cm},clip]{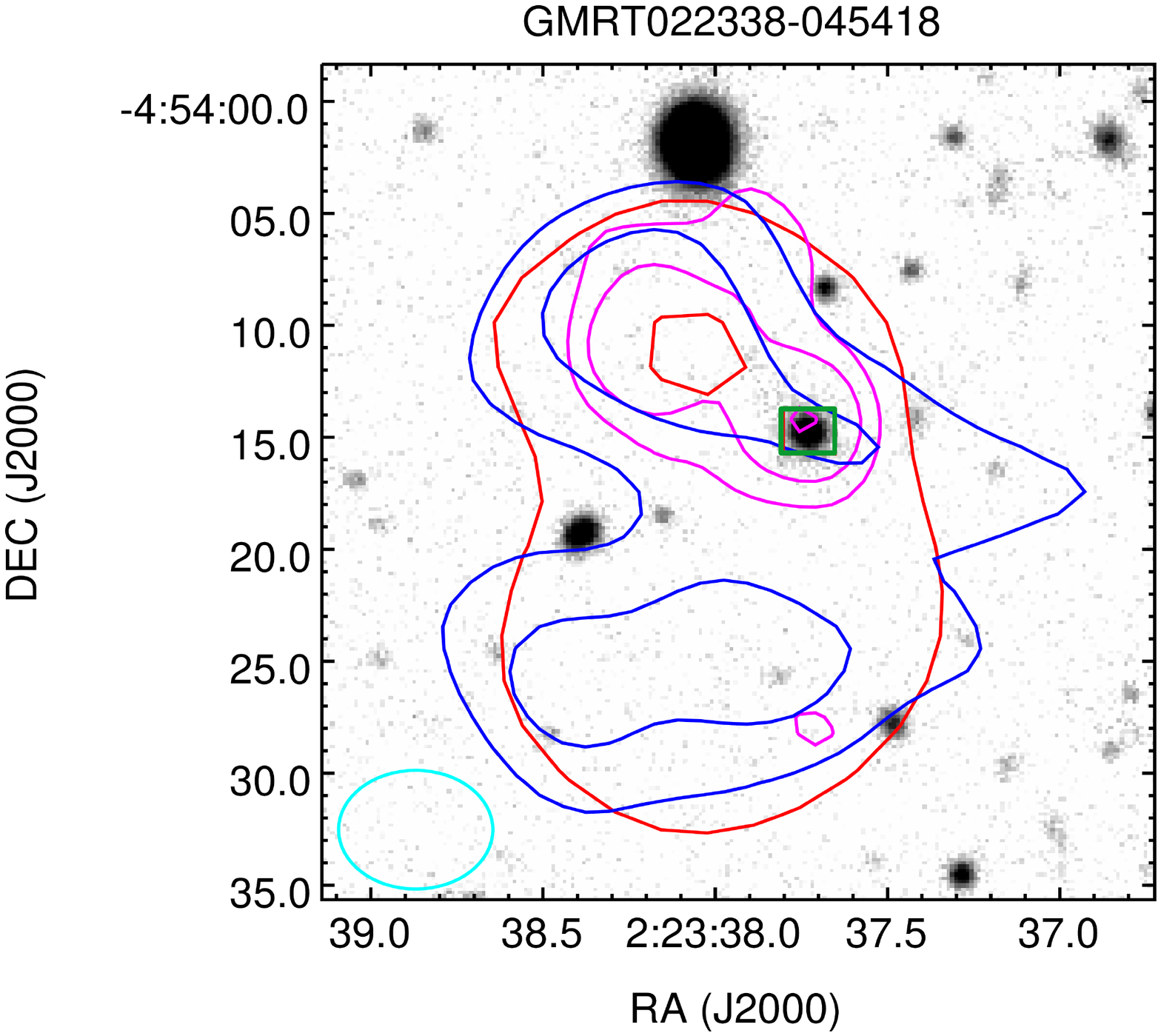}
\includegraphics[angle=0,width=9.0cm,trim={0.0cm 0.0cm 0.0cm 0.0cm},clip]{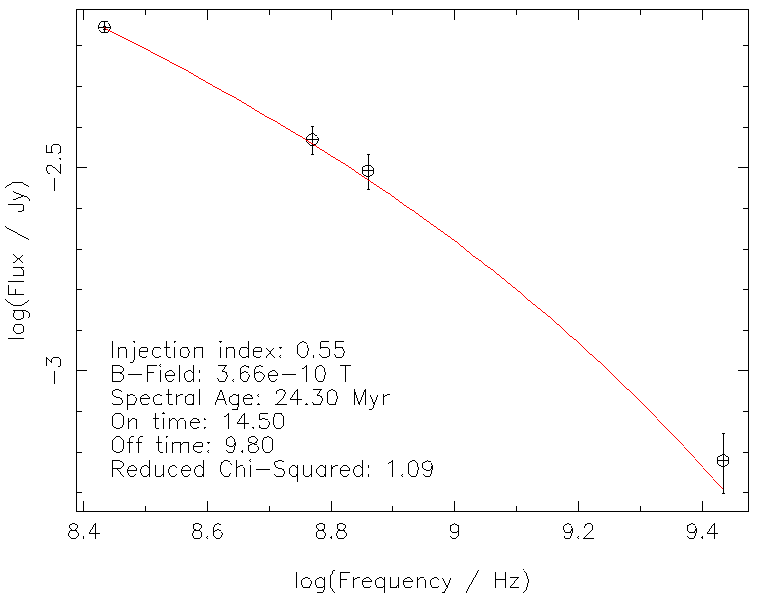}
\includegraphics[angle=0,width=8.0cm,trim={0.0cm 4.0cm 0.0cm 6.5cm},clip]{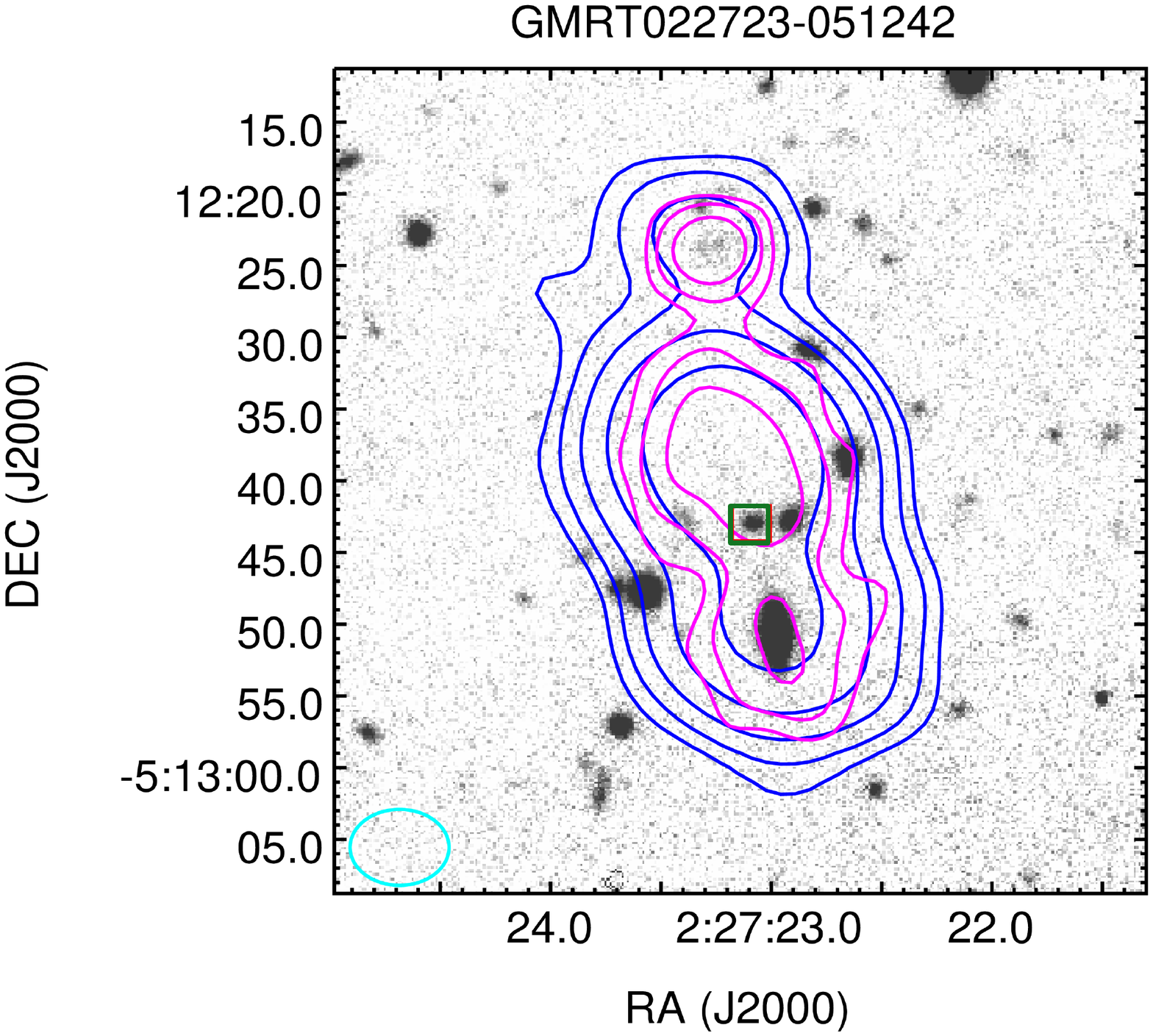}
\includegraphics[angle=0,width=9.0cm,trim={0.0cm 0.0cm 0.0cm 0.0cm},clip]{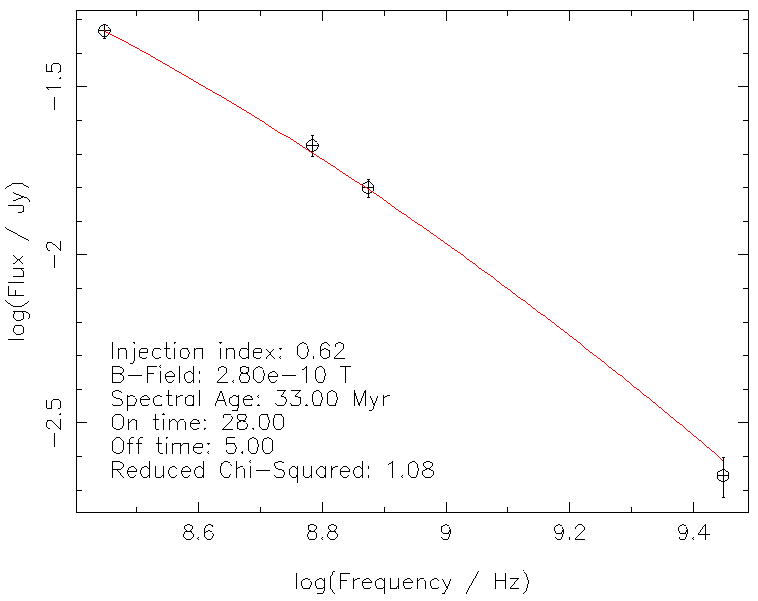}
\caption{{\it Left panel}: The images of remnant sources. The radio contours of band-3 uGMRT (in Blue), 325 MHz (in Red), and 1.5 GHz JVLA (in Magenta) are overplotted on the corresponding HSC $i$ band optical image. The radio contours are at 3$\sigma$ $\times$ (1, 2, 4, 8, 16, .....) levels and the optical image is logarithmically scaled. The uGMRT band-3 synthesized beam of 6$^{\prime\prime}{.}$7 $\times$ 5$^{\prime\prime}{.}$3 is shown in the bottom left corner. The potential host galaxy of a remnant is marked with a green box around it. {\it Right panel}: The best fit radio SEDs of our remnant sources. The solid red curve represents model fitted to the data points. }
\label{fig:ImSED}
\end{figure*}
\addtocounter{figure}{-1}
%
%
\begin{figure*}[hp]
\includegraphics[angle=0,width=9.0cm,trim={0.0cm 4.0cm 0.0cm 6.5cm},clip]{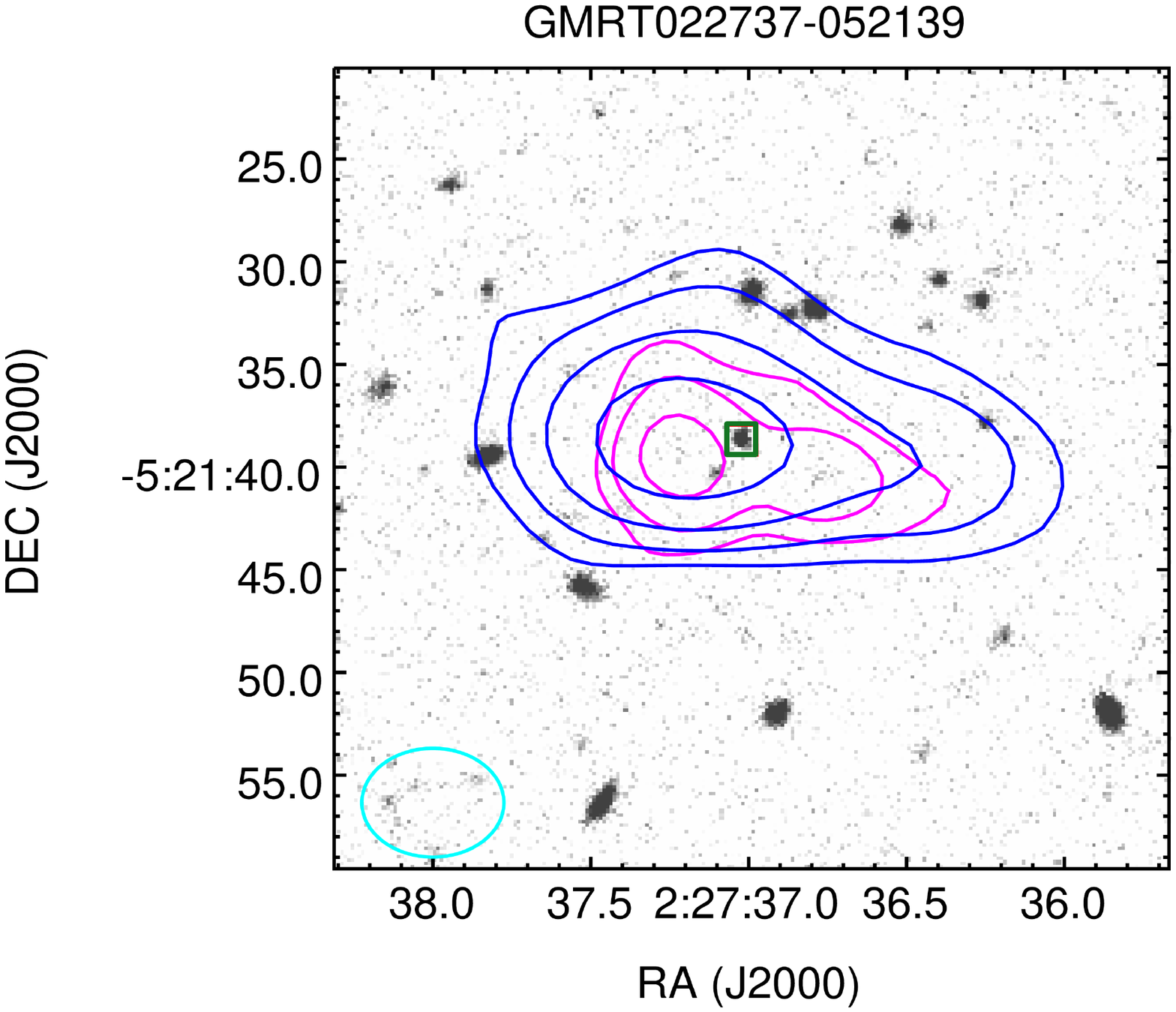}
\includegraphics[angle=0,width=9.0cm,trim={0.0cm 0.0cm 0.0cm 0.0cm},clip]{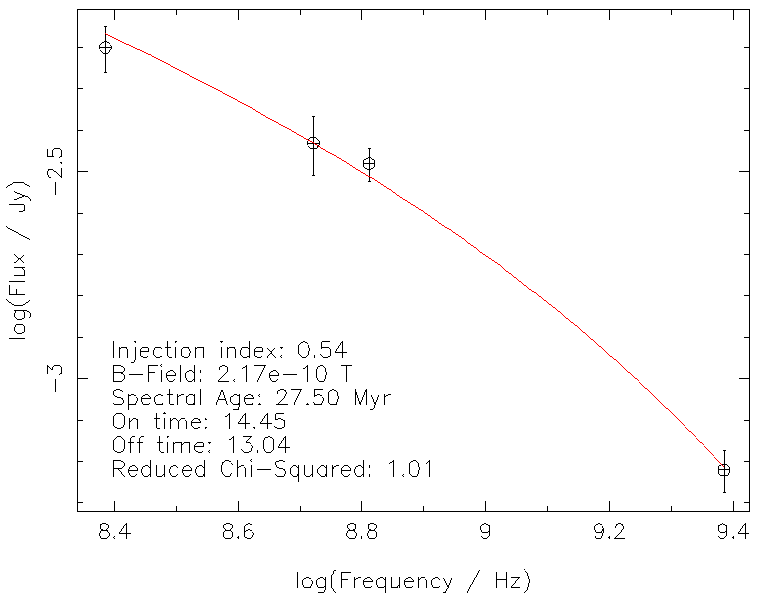}
\includegraphics[angle=0,width=9.0cm,trim={0.0cm 4.0cm 0.0cm 6.5cm},clip]{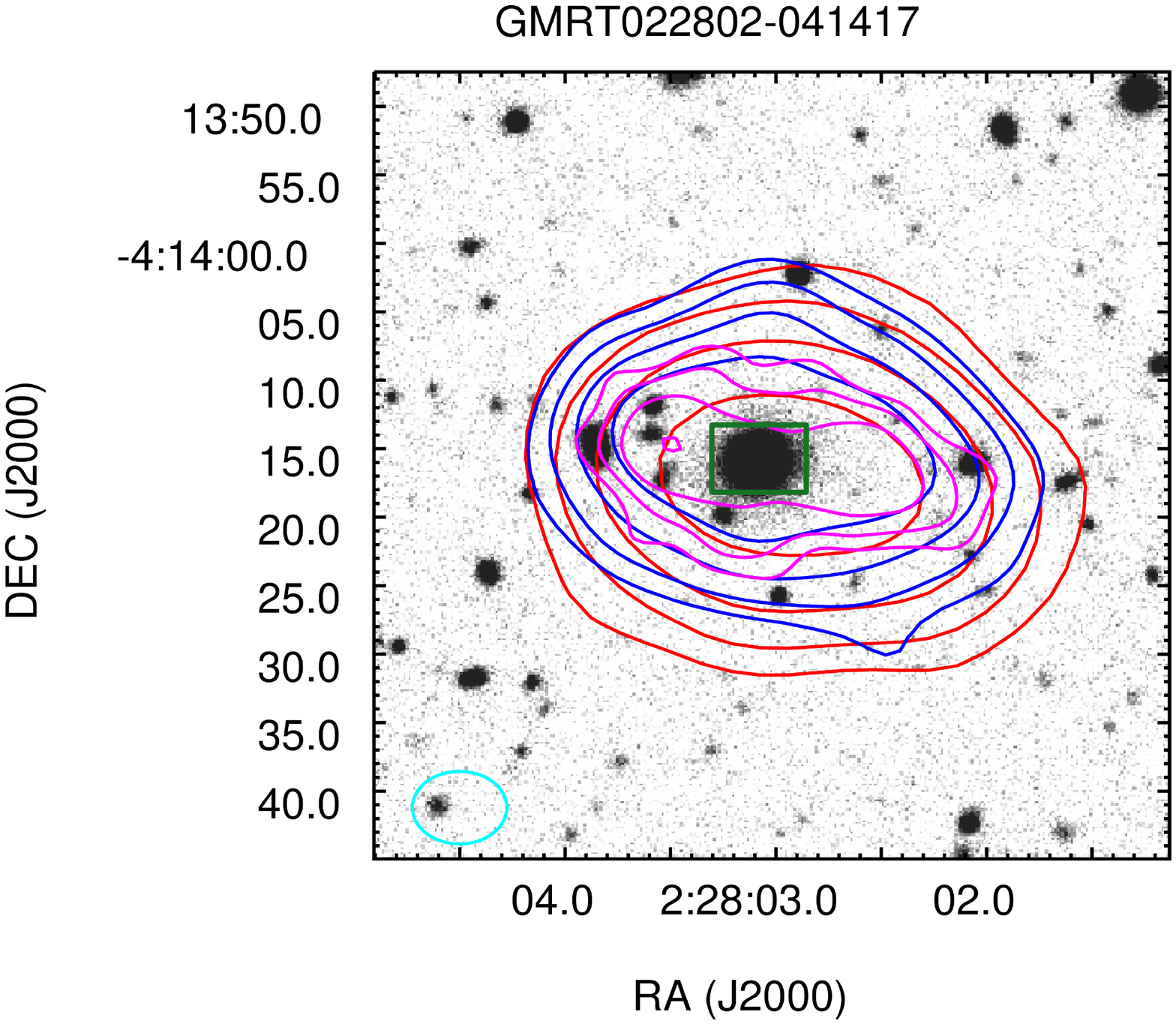}
\includegraphics[angle=0,width=9.0cm,trim={0.0cm 0.0cm 0.0cm 0.0cm},clip]{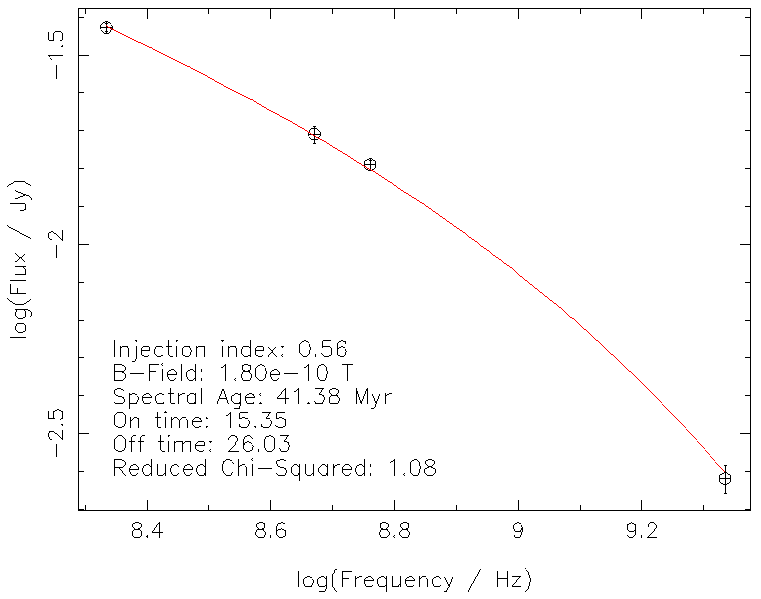}
\caption{\it - continue}
\label{fig:ImSED}
\end{figure*}
%
%
%
We note that our remnants show ${\nu}_{\rm b,~low}$ in the range of 250 MHz to 875 MHz above 
which their spectra steepen. The values of ${\nu}_{\rm b,~low}$ are consistent with the fact that our sources exhibit strong 
spectral curvature between 150 MHz and 1.5 GHz, {\ie}${\alpha}_{\rm 150~MHz}^{\rm 325~MHz}$ - 
${\alpha}_{\rm 400~MHz}^{\rm 1.5~GHz}$ $\geq$0.5. Although, in all sources, 
the ${\nu}_{\rm b,~high}$ above which spectrum falls off exponentially, lies outside the frequency coverage 
($>$1.5 GHz) (see Table~\ref{tab:energetics}), and it is determined from the SED modelling. 
The knowledge of break frequencies allows us to obtain spectral age estimates.   
Based on the best spectral fits, we find the total spectral ages ($t_{\rm s}$) of our remnants 
in the range of 20.3 Myrs to 41.4 Myrs. 
The duration of the active phase ($t_{\rm ON}$) is in the range 14.5 Myr to 28.0 Myr, while sources have spent nearly 
5.0 Myr to 26.0 Myr in the remnant phase (see Table~\ref{tab:energetics}). 
We note that the typical errors associated with the spectral ages 
($t_{\rm s}$, $t_{\rm ON}$ and $t_{\rm OFF}$) are nearly 10 per cent. 
Although, the quoted errors are rendered by the model fittings and true uncertainties can be much 
larger owing to the several assumptions made in computing spectral ages \citep{Harwood17}. 
Hence, we caution that the spectral age estimates should only be treated as characteristic timescales.   
\par
Further, we compute the average speed of lobes along the jet axis by dividing the average distance between centre and the outer 
edge of lobes with the timescale of active phase $t_{\rm ON}$. We assume that our sources are lying in the plane of sky, {\ie}the angle between the jet axis and the line-of-sight is 90$^{\circ}$. 
For our remnants, we find that the average speeds of lobes are in the range of 0.01$c$ to 0.07$c$, which 
are similar to those found for powerful radio galaxies \citep[see][]{ODea09}. In other words, dynamical ages, if derived using 
a typical average speed of lobes, would be consistent with the spectral ages. 
We also examine the ratio of remnant age to total source age ($t_{\rm OFF}$/$t_{\rm s}$) which describes the fraction 
of total life a source spent in the remnant phase. For our remnant sources, we find $t_{\rm OFF}$/$t_{\rm s}$ in the 
range of 0.16 to 0.63 (see~Table~\ref{tab:energetics}). The wide range of $t_{\rm OFF}$/$t_{\rm s}$ infers that 
our sources are in different phases of their evolution. One of the noteworthy examples in our sample is 
GMRT022802-041417, which has spent nearly 63 per cent of its total life in the remnant phase. 
The active phase duration of this source is only 15.4 Myr, while it has spent nearly 26.0 Myr in the remnant phase. 
Thus, we emphasize that despite the limited frequency coverage (150 MHz $-$ 1.5 GHz) our observations are capable of detecting remnants in the different phases. 
\subsection{Comparison with other remnants}
\label{sub:comparison}
\begin{figure}[ht]
\includegraphics[angle=0,width=9.0cm,trim={0.1cm 6.5cm 0.0cm 7.5cm},clip]{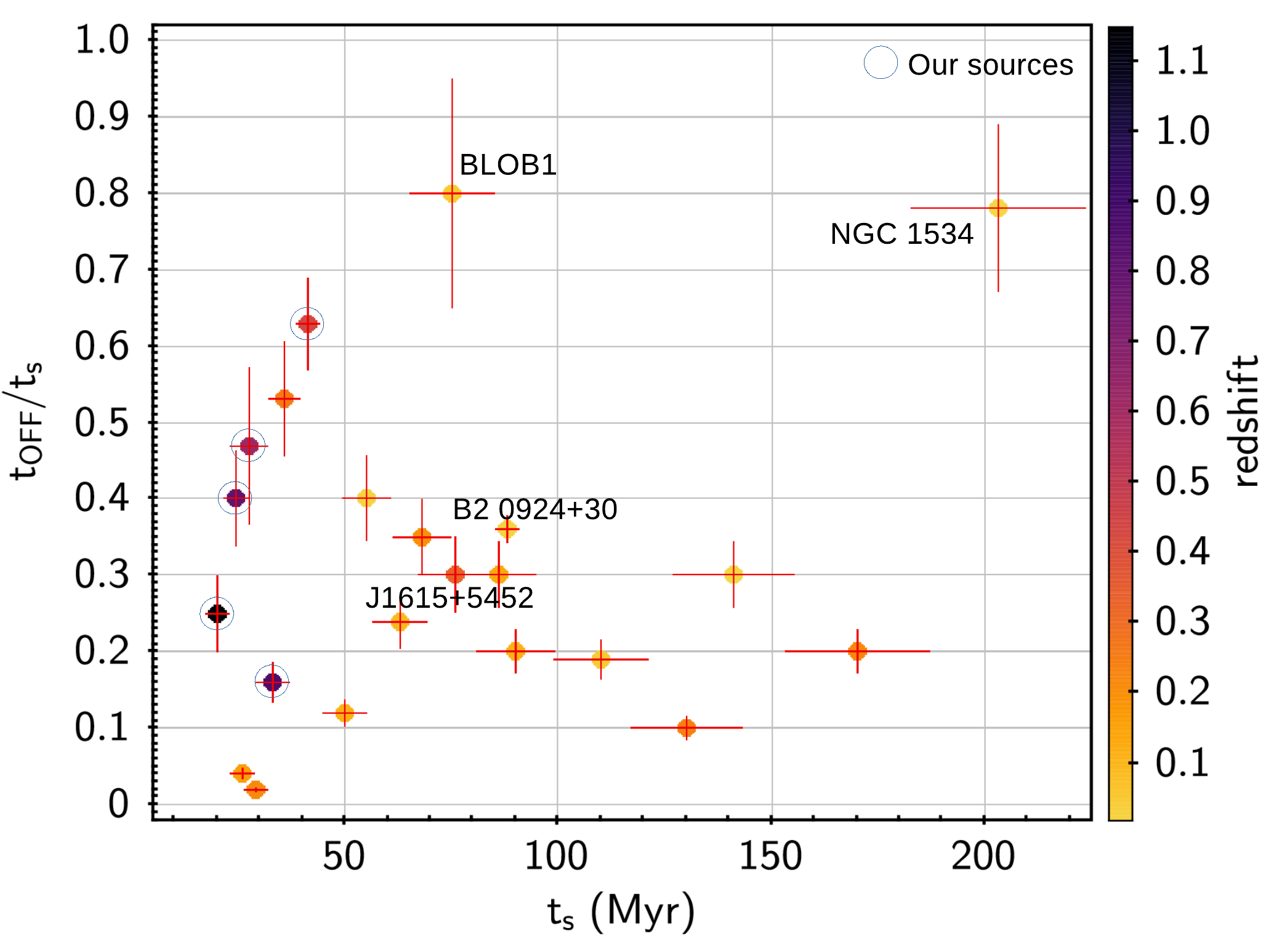}
\caption{The plot of total source age $t_{\rm s}$ versus fractional remnant time-scale ($t_{\rm OFF}$/$t_{\rm s}$). 
The vertical colour bar indicates the redshifts of remnant hosts. The individually studied remnant 
sources are marked by their popular names. Our sample sources are marked by large Blue circles.}
\label{fig:Age}
\end{figure}

To understand the nature of our remnants, we compare their spectral ages with the previously studied remnant sources.
The comparison sources include BLOB1 \citep{Brienza16}, B2 0924+30 \citep{Shulevski17}, J1615+5452 \citep{Randriamanakoto20}, 
NGC 1534 \citep{Duchesne19}, four sources from \cite{Murgia11} and nine sources from \cite{Parma07}. 
In Figure~\ref{fig:Age}, we show a plot of total source age ($t_{\rm s}$) versus fractional remnant 
duration ($t_{\rm OFF}$/$t_{\rm s}$). 
It is evident that our sample remnants have systematically lower spectral ages than that for most of the remnant sources. 
There are only three sources, in addition to our five sample sources that have spectral ages $<$50 Myr. 
However, we note that a simple comparison of the spectral ages in a heterogeneous sample of remnants 
may not yield meaningful insights considering the fact that source age depends on various factors such as physical radio size, magnetic field, large-scale environment and redshift \citep{Turner18}. 
Therefore, we consider the effects of these parameters in our further discussion. 
\par
We note that \cite{Murgia11} found a longer duration of source ages ($t_{\rm s}$ $>$50 Myr) 
for their remnant sources residing in cluster environments. 
The longer age for sources residing in clusters can be attributed to the slow or arrested expansion of 
lobes due to the high pressure of the dense intra-cluster medium. 
Hence, fading lobes of a remnant can last longer if energy losses due to dynamical effects are insignificant or absent \citep{Murgia11}. 
We point out that most of the remnant sources presented in \cite{Parma07} also reside in cluster environments, and hence, 
they show high spectral ages. There are only three sources in the sample of \cite{Parma07} with $t_{s}$ $<$50 Myr. 
Two of these sources with small values of $t_{\rm OFF}$/$t_{\rm s}$ are recently switched-off sources and show 
unusually small radio sizes ($<$10 kpc). 
The relatively short source ages can be understood as jets and lobes taking less time to grow 
to the smaller sizes. 
The radio sizes of our remnants, in the range of 190 kpc to 643 kpc, are similar to other remnants that show higher 
spectral ages. We point out that all our remnants with lower ages reside in non-cluster environments, while sources of similar 
radio sizes but higher spectral ages reside mostly in cluster environments. Hence, as expected,  
our sample sources residing in less dense environments fade away more rapidly due to the faster expansion of lobes. 
Thus, the relatively low spectral ages of our remnants, when compared to the remnants residing in cluster environments,  can be partly attributed to the less dense large-scale environment.  
\par
Further, remnants with large radio sizes are expected to show high spectral ages owing to the fact that jets and lobes take a longer time to traverse longer distances and to fill up large volumes. 
Both BLOB1 and NGC~1534 are the examples of old large-size remnants with their 
radio sizes spanning up to nearly 700 kpc and 600 kpc, respectively \citep[see][]{Brienza16,Duchesne19}. 
The spectral ages of BLOB1 and NGC~1534 are 75 Myr and 203 Myr, respectively. 
The high value $t_{\rm OFF}$/$t_{\rm s}$ $\sim$ 0.8 confirms them to be old remnants that have spent nearly 80 per cent of 
their life in the remnant phase. Therefore, high spectral ages of BLOB1 and NGC~1534 remnants can 
be attributed to the combined effects of large radio size and long duration of the remnant phase. 
We note that one of our sample sources GMRT022318-044526 has the radio size of 
643 kpc, but it has source age ($t_{\rm s}$) of only 20.3 Myr. 
The low value of $t_{\rm OFF}$/$t_{\rm s}$ = 0.25 
suggests it to be a relatively younger remnant. Notably, GMRT022318-044526 reside at a much higher 
redshift ($z$) = 1.15 (see~Table~\ref{tab:sample}).  
We would like to emphasize that all our sample sources are at relatively higher redshifts ($z$ $\sim$ 0.44$-$1.15) 
which is evident from Figure~\ref{fig:Age}. At high redshifts, inverse Compton losses dominate owing to the high density 
of Cosmic Microwave Background (CMB) photons. The inverse Compton equivalent magnetic field ($B_{\rm CMB}$) scales 
with $(1+z)^{2}$ (see equation~\ref{eq:1}). Thus, due to increased inverse Compton losses, remnant sources at higher 
redshifts are expected to fade away much faster than their counterparts at lower redshifts. 
Hence, relatively lower spectral ages of our remnants can be largely attributed to their high redshifts. 
\par        
We note that, to nullify the effects of the environment, radio size and redshift, one should compare remnants 
matched in these parameters. However, due to the paucity of remnant sources, we do not have such a comparison sample. 
It is worth mentioning that one remnant source WNB1127.5+4927 in the sample of \cite{Parma07} is 
reported to reside in non-cluster environment at redshift $z= 0.25$ with its radio size of 211 kpc.   
The WNB1127.5+4927 shows similarity with our sample sources in terms of environment, size, and redshift, and hence, as expected,  
it has $t_{s}$ = 36 Myr and $t_{\rm OFF}$ = 19 Myr, similar to our remnants.       
The remnant source J1615+5452 residing in non-cluster environments at $z$ = 0.33 and radio size of 100 kpc, 
is also similar to our remnants, but it is reported to have higher spectral age 
with $t_{s}$ = 76 Myr and $t_{\rm OFF}$ = 22 Myr \citep[see][]{Randriamanakoto20}. 
A careful inspection of the spectral fitting of this source reveals that a statistically 
better fit with reduced ${\chi}^{2}$ = 1.42 yields $t_{s}$ = 35.9 Myr and $t_{\rm OFF}$ = 21.5 Myr, although it requires 
unusually flat ${\alpha}_{\rm inj}$ = 0.4 and a high value of $B_{\rm eq}$ = 12 $\mu$G. 
We emphasize that a more robust spectral modelling using flux densities measured across a wide range of frequencies is required.  
In this context, the role of deep multi-frequency radio surveys is discussed in the next Section.  
\section{Potential of deep radio continuum surveys from the SKA and its pathfinders}
\label{sec:SKA}
Our full synthesis band-3 uGMRT observations have provided sensitive images with noise-rms down 
to 30~${\mu}$Jy~beam$^{-1}$ and 
angular resolution of 6$^{\prime\prime}$.7 $\times$ 5$^{\prime\prime}$.3. 
Owing to its depth and angular resolution, we discovered wing-shaped radio morphology in one of our sample sources 
GMRT022338-045418. The diffuse emission of low-surface-brightness related to the southern component of this source 
is completely missed in the 1.5 GHz JVLA observations. Previous less sensitive 325 MHz GMRT observations with somewhat coarser 
resolution (10$^{\prime\prime}$.2 $\times$ 7$^{\prime\prime}$.9) could not decipher radio structures in 
GMRT022338-045418 (see~Figure~\ref{fig:ImSED}). 
Therefore, in the context of ongoing and planned deep radio continuum surveys, our band-3 uGMRT observations can be considered as a 
test-bed for discovering remnants possessing diffuse low-surface-brightness emission. 
For instance, the Evolutionary Map of the Universe (EMU) survey conducted with the Australian Square Kilometer Array Pathfinder 
(ASKAP) at 944 MHz provides deep continuum images with noise-rms of 25$-$30 $\mu$Jy beam$^{-1}$ and resolution of 
11$^{\prime\prime}$ $-$ 18$^{\prime\prime}$ over a large area of 270 deg$^{2}$ \citep{Norris21}. Despite being at 
a relatively higher frequency, EMU is capable of discovering large-scale diffuse emission owing to the short baselines up to 22m available in the ASKAP. 
\par
Recently, in a small area of 8.3 deg$^{2}$ of GAMA-23 field \cite{Quici21} identified 
ten remnant candidates by using 119$-$216 MHz Murchison Wide-field Array (MWA) observations, 400 MHz uGMRT observations, 
944 MHz EMU survey, and 5.5~GHz Australian Telescope of Compact Array (ATCA) observations. 
The multi-frequency observations were used to build broad-band SEDs and confirm the remnant status of radio galaxies. 
We emphasize that recent studies indicate remnant sources to be more abundant than that thought earlier. 
For instance, \cite{Singh21} used the spectral curvature criterion and identified a large sample 
of 48 small-size ($<$30$^{\prime\prime}$) remnant candidates in 12.5 deg$^{2}$ area in 
the {\em XMM-LSS} field. The majority of small-size remnants are found at higher redshift ($z$ $>$1.0). 
In the {\em XMM-LSS} field, we also identified 21 large-size ($>$30$^{\prime\prime}$) remnant candidates which show 
distinct double-lobe morphology with absent core and tend to be more powerful 
(L$_{\rm 150~MHz}$ $>$ 10$^{25}$ W~Hz$^{-1}$) radio sources (Dutta et al. 2022, {\it under review}). 
Thus, remnant searches performed in the deep fields have demonstrated the existence of a variety of remnant sources residing 
across a wide range of redshifts. Although, most of the searches in the deep field are limited to only a small sky-area of a few deg$^{2}$ and rendered only small samples. 
The large-area deep surveys have the potential to discover a large population of remnants that 
would offer better insights into the evolution of radio galaxies in the remnant phase. 
\par
In Section~\ref{sec:energetic}, we pointed out that the radio observations over a wide range 
of frequencies are needed for robust determination of spectral ages of remnants. 
The upcoming radio continuum surveys from the SKA telescope covering a wide range of frequencies, 
{\ie}50 MHz $-$ 350 MHz with SKA1$-$low, and 350 MHz $-$ 15.3 GHz with SKA1-mid would be ideal for spectral modelling. 
As per the baseline design performance, SKA1-mid is expected to provide noise-rms of 4.4 $\mu$Jy beam$^{-1}$ 
in one hour observing time in the 0.35$-$1.05 GHz band centered at 770 MHz 
(see SKA factsheet\footnote{https://www.skatelescope.org/wpcontent/uploads/2018/08/16231-factsheet-telescopes-v71.pdf}).
Thus, the large-area multi-frequency deep radio surveys from the SKA and its pathfinder are expected to unveil a large population of remnants distributed over different evolutionary phases. 
\section{Conclusions}
\label{sec:conclusions}
In this paper, we presented the new band-3 uGMRT observations of five remnant sources selected based on 
the strong spectral curvature, emission of low-surface-brightness from lobes, and absent radio core in the 3.0 GHz VLASS images. 
We performed modelling of radio SEDs using observations from the LOFAR at 144 MHz, GMRT at 325 MHz, uGMRT at 400 MHz (band-3) 
and JVLA at 1.5~GHz. The salient results of our study are outlined as below.

\begin{itemize}
\item With our deep full synthesis band-3 uGMRT observations (median noise-rms $\sim$ 30~mJy~beam$^{-2}$) 
we confirm the presence of low-surface-brightness emission in all 
our five remnant sources. The surface brightness at 400 MHz is found in the range of 36 mJy~arcmin$^{-2}$ to 201 
mJy~arcmin$^{-2}$, which is similar to those found in remnants studied in the literature \citep{Brienza17}.     
\item With our band-3 uGMRT observations, we discovered wing-shaped radio morphology in one of our sample sources GMRT022338-045418. 
Previous 325 MHz GMRT observations were unable to reveal structures due to its coarser resolution, while 1.5~GHz JVLA 
observations did not detect diffuse emission from the southern component. Both northern as well as southern components show two peaks that are similar to the primary and secondary lobes commonly seen in the Z-shaped and X-shaped 
radio sources. 
Despite its extended morphology, this source is faint with the total flux density of only 3.1$\pm$0.5 mJy at 400 MHz, which in 
turn yields low surface-brightness of 36~mJy~arcmin$^{-2}$. 
\item We fitted the radio SEDs 
with CI$_{\rm OFF}$ model that assumes a continuous injection phase for the duration of $t_{\rm ON}$, and thereafter, a 
remnant phase for the duration of $t_{\rm OFF}$ with total source age $t_{\rm s}$ = $t_{\rm ON}$ + $t_{\rm OFF}$. 
For our remnant sources, the best fit models yield source ages ($t_{\rm s}$) in the range of 20.3 Myr to 41.4 Myr.       
\item We find that our small sample of remnants shows a wide distribution of $t_{\rm OFF}$/$t_{\rm s}$ in the range of 0.16 to 0.63. 
In other words, our sample sources have spent 16 per cent to 63 per cent of their total life in the remnant phase. Hence, 
our sources belong to the different phases of evolution. 
\item The estimated spectral ages of our remnants are, in general, lower than those found for several remnants previously studied in 
the literature. We caution that spectral age depends on various factors such as the radio size, magnetic field, redshift, 
and the large-scale environment, so we need to account for these parameters while comparing spectral ages. Notably, 
remnants residing in cluster environments show higher spectral ages due to less or insignificant dynamical energy losses 
from the reduced or absent expansion of lobes embedded within the dense intra-cluster medium. Further, we found that our remnant 
sources have systematically higher redshifts than most of the sources reported in the literature. Therefore, 
inverse Compton energy losses are more dominant for our high redshift sources, which in turn reduces their source ages. 
\item We note that the total source ages of our remnants are similar to one of the sample sources WNB1127.5+4927 of \cite{Parma07}, 
which has similar radio size, redshift, and environment. 
\item Our study, limited to a small sample lying within a small area of 2.3 deg$^{2}$ covered with one pointing of  band-3 uGMRT observations, demonstrates the potential of deep large-area surveys from the SKA and its pathfinders. 
Based on our study, we expect a large population of remnants to be detected with the ongoing and planned deep 
radio continuum radio surveys.
\end{itemize}
%
%
%
%
%
\section*{Acknowledgements}
We thank the anonymous reviewer for carefully reading the manuscript and giving valuable comments that improved the manuscript. SD, VS and AK acknowledge the support from the Physical Research Laboratory, Ahmedabad, funded by the Department of Space, Government of India. 
CHI and YW acknowledge the support of the Department of Atomic Energy, Government of India, under project no. 12-R\&D-TFR5.02-0700.
We thank the staff of GMRT who have made these observations possible. GMRT is run by the National Centre for Radio 
Astrophysics of the Tata Institute of Fundamental Research.
This paper is based on data collected from the Subaru Telescope and retrieved from the HSC data archive system, which is operated by the Subaru Telescope and Astronomy Data Center (ADC) at NAOJ. Data analysis was in part carried out with the cooperation of Center for Computational Astrophysics (CfCA), NAOJ. 
\vspace{-1em}
%
\bibliography{JaaRemnantsRevised}{}
%
\nocite{*}
\end{document}